\documentclass[%
 reprint,
 nolongbibliography,
 amsmath,amssymb,
 aps,
 prmaterials
]{revtex4-2}

\usepackage{graphicx}
\usepackage{dcolumn}
\usepackage{bm}
\usepackage{hyperref}
\usepackage{xcolor}
\usepackage{natbib}

\begin{document}


\title{Dramatic elastic response at the critical end point in UTe$_2$}

\author{Michal Vališka}
 \email{michal.valiska@matfyz.cuni.cz}
 \affiliation{%
 Charles University, Faculty of Mathematics and Physics, Department of Condensed Matter Physics, Ke Karlovu 5, 121 16 Prague 2, Czech Republic}%
 
\author{Tetiana Haidamak}%
 \affiliation{%
 Charles University, Faculty of Mathematics and Physics, Department of Condensed Matter Physics, Ke Karlovu 5, 121 16 Prague 2, Czech Republic}%

\author{Andrej Cabala}%
 \affiliation{%
 Charles University, Faculty of Mathematics and Physics, Department of Condensed Matter Physics, Ke Karlovu 5, 121 16 Prague 2, Czech Republic}%

 \author{Jiří Pospíšil}%
 \affiliation{%
 Charles University, Faculty of Mathematics and Physics, Department of Condensed Matter Physics, Ke Karlovu 5, 121 16 Prague 2, Czech Republic}%

 \author{Gaël Bastien}%
 \affiliation{%
 Charles University, Faculty of Mathematics and Physics, Department of Condensed Matter Physics, Ke Karlovu 5, 121 16 Prague 2, Czech Republic}%

 \author{Tatsuya Yanagisawa}%
 \affiliation{%
Department of Physics, Hokkaido University, Sapporo 060-0810, Japan}%

 \author{Petr Opletal}%
 \affiliation{Advanced Science Research Center, Japan Atomic Energy Agency, Tokai, Ibaraki 319-1195, Japan}%

 \author{Hironori Sakai}%
 \affiliation{Advanced Science Research Center, Japan Atomic Energy Agency, Tokai, Ibaraki 319-1195, Japan}%

 \author{Yoshinori Haga}%
 \affiliation{Advanced Science Research Center, Japan Atomic Energy Agency, Tokai, Ibaraki 319-1195, Japan}%

 \author{Atsuhiko Miyata}%
 \affiliation{Hochfeld-Magnetlabor Dresden (HLD-EMFL), Helmholtz-Zentrum Dresden-Rossendorf, 01328 Dresden, Germany}%

 \author{Denis Gorbunov}%
 \affiliation{Hochfeld-Magnetlabor Dresden (HLD-EMFL), Helmholtz-Zentrum Dresden-Rossendorf, 01328 Dresden, Germany}%

 \author{Sergei Zherlitsyn}%
 \affiliation{Hochfeld-Magnetlabor Dresden (HLD-EMFL), Helmholtz-Zentrum Dresden-Rossendorf, 01328 Dresden, Germany}%

  \author{Vladimír Sechovský}%
 \affiliation{%
 Charles University, Faculty of Mathematics and Physics, Department of Condensed Matter Physics, Ke Karlovu 5, 121 16 Prague 2, Czech Republic}%

  \author{Jan Prokleška}%
 \affiliation{%
 Charles University, Faculty of Mathematics and Physics, Department of Condensed Matter Physics, Ke Karlovu 5, 121 16 Prague 2, Czech Republic}%

\date{\today}

\begin{abstract}
The first-order transition line in the \textit{H-T} phase diagram of itinerant electron metamagnets terminates at the critical end point—analogous to the critical point on the gas-liquid condensation line in the \textit{p-T} phase diagram. To unravel the impact of critical magnetic fluctuations on the crystal lattice of a metamagnet at the critical end point, we performed an ultrasonic study of the itinerant electron metamagnet UTe$_2$ across varying temperatures and magnetic fields. At temperatures exceeding 9 K, a distinct V-shaped anomaly emerges, precisely centered at the critical field of the metamagnetic transition in the isothermal field dependence of elastic constants. This anomaly arises from lattice instability, triggered by critical magnetic fluctuations via strong magnetoelastic interactions. Remarkably, this effect is maximized precisely at the critical-end-point temperature. Comparative measurements of another itinerant metamagnet, UCoAl, reveal intriguing commonalities. Despite significant differences in the paramagnetic ground state, lattice symmetry, and the expected metamagnetic transition process between UTe$_2$ and UCoAl, both exhibit similar anomalies in elastic properties near the critical end point. These shared aspects may hold universality for other itinerant electron metamagnets. 

\end{abstract}

\maketitle


\section{\label{sec:Intro}Introduction}

The ground state of itinerant electron metamagnets (IEMs) is paramagnetic. In weak magnetic fields, they behave as enhanced paramagnets. A notable feature of IEMs is the abrupt increase in magnetization when a critical magnetic field, denoted as $H_c$, is applied at low temperatures, due to a first-order metamagnetic transition (MT). The critical field, $\mu_0H_c$, varies across different IEMs, ranging from approximately 0.8~T in UCoAl\cite{HAVELA1997} to about 70~T in YCo$_2$\cite{GOTO1989,GOTO1990}. Beyond $H_c$, the magnetic moments align with the applied magnetic field, resembling a collinear ferromagnet yet maintaining a paramagnetic state in what is termed the polarized paramagnetic (PPM) regime\cite{Knafo2021}.

UCoAl is one of the most studied IEMs. Its magnetism is governed by $5f$ electron magnetic moments of U ions. It exhibits strong uniaxial anisotropy causing its Ising character. The MT in UCoAl is triggered by a low magnetic field but only when aligned along the hexagonal $c$-axis, whereas its magnetization response to magnetic fields applied in perpendicular directions up to 35 T resembles a Pauli paramagnet\cite{SECHOVSKY1986}. 

UTe$_2$ has garnered attention for its recently discovered unconventional superconductivity\cite{Ran2019,Aoki_2022}. The physics of UTe$_2$ is often discussed in connection with ferromagnetic superconductors URhGe\cite{Hardy2011,Levy2005}, and UCoGe\cite{HUY2009,Aoki2009,Knafo2019}, which share orthorhombic structures and exhibit similar anisotropy in magnetic-field-induced phenomena. Unlike ferromagnetic URhGe and UCoGe, UTe$_2$ possesses a paramagnetic ground state\cite{Paulsen2021} with antiferomagnetic fluctuations\cite{Knafo2021a}. At low temperatures, it undergoes an MT exclusively when the field is applied along the $b$-axis at $\mu_0H_c\approx$ 35~T\cite{Miyake2019}, placing UTe$_2$ among the anisotropic uranium IEMs alongside UCoAl. This similarity has motivated our comparative study of the metamagnetism in these two compounds.

The MT in IEMs is accompanied by pronounced changes in physical properties coupled with the magnetic state. The lattice parameters change suddenly across MT due to strong magnetoelastic coupling. The hexagonal UCoAl lattice contracts along the field direction ($c$-axis) and expands in perpendicular directions so that the volume increases\cite{PROKLESKA2007,HONDA2000} that can be understood in terms of a partial localization of U $5f$-electron states. The orthorhombic UTe$_2$ lattice contracts along the field direction ($b$-axis), expands along $c$, and remains almost intact along $a$. That leads to a reduction of the lattice volume\cite{Miyake2022} at the MT.  The volume shrinkage in UTe$_2$ at $H_c$, contrary to the volume expansion in typical heavy fermion metamagnets may be due to a specific electronic structure transition related to the dual nature of uranium $5f$-electrons.

Metamagnetic transitions also significantly impact magnetotransport properties, reflecting changes in the electronic structure. The electrical resistivity of UCoAl shows at MT a positive (negative) step for the current parallel (perpendicular) to the $c$-axis\cite{Matsuda2000}. The ordinary Hall and Seebeck coefficients change abruptly at $H_c$ at the lowest temperature evidencing a change in the density of states at the Fermi level due to the MT\cite{Matsuda2000,Matsuda1999,Combier2013,Palacio2013}. The resistivity of UTe$_2$ for $a$-axis current exhibits a sharp positive step at MT\cite{Knafo2019a}. The field-dependent discontinuities of the ordinary Hall and Seebeck coefficients at MT are indicative of changes in the Fermi surface due to MT\cite{Niu2020}.

\begin{figure}
    \includegraphics[width=0.48\textwidth]{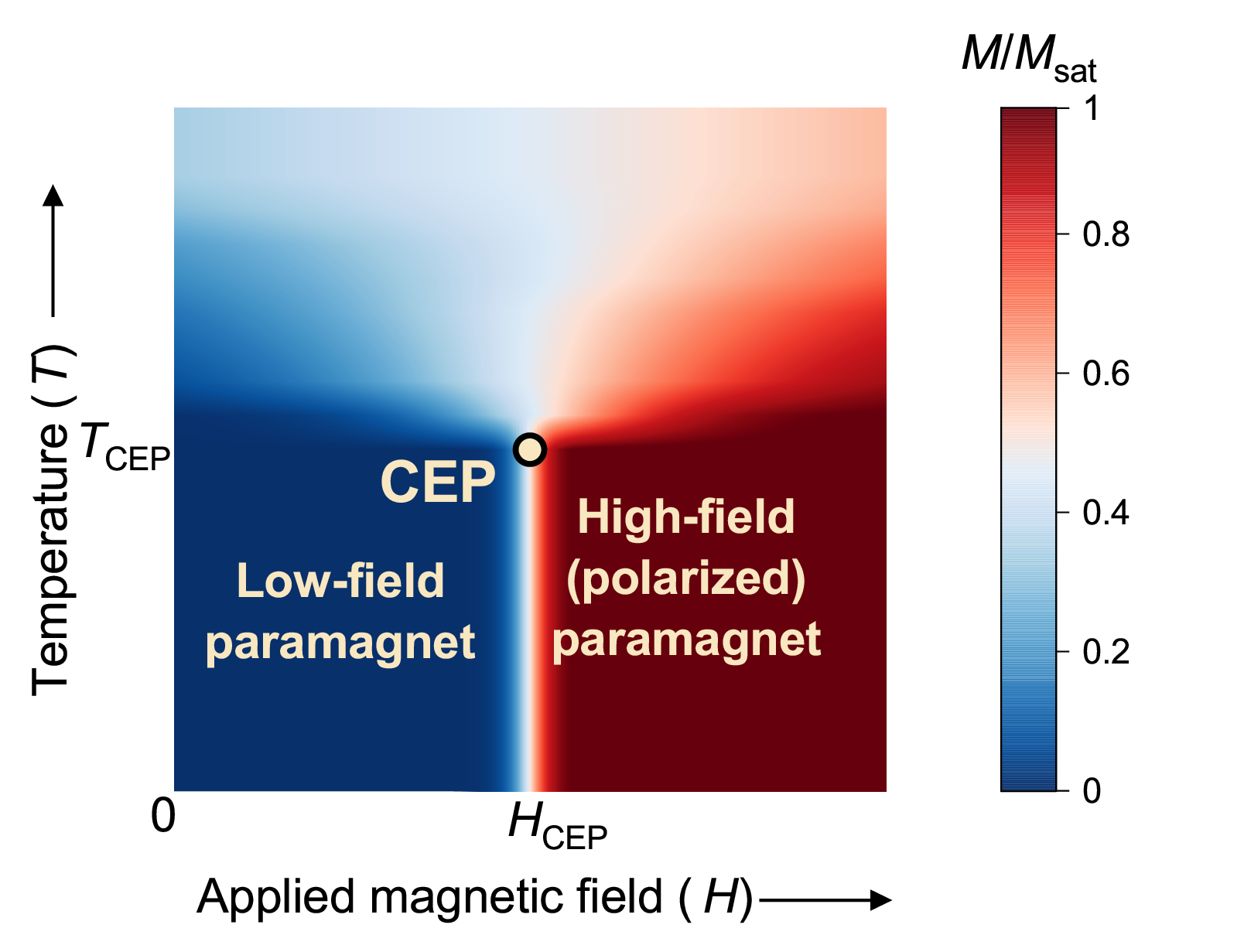} 
    \caption{\label{fig1}Schematic $H-T$ phase diagram of an itinerant electron metamagnet.} 
\end{figure}

The metamagnetic phase transition line in the $H-T$ phase diagram of itinerant electron metamagnets (see Fig.~\ref{fig1}) terminates at a critical end point\cite{Millis2002,Belitz1999} (CEP) having coordinates ($H_{CEP}$, $T_{CEP}$). Beyond CEP no phase transition is observed. The CEP in the $H-T$ phase space of IEMs serves as an analogy to the critical point (CP) in the gas-liquid $p-T$ phase diagram\cite{Fisher1990,Stishov2020}. The existence of CEP was first indicated in UCoAl at $\mu_0H_{CEP}$ somewhat smaller than 1~T and $T_{CEP}$ $\approx$ 11 $\pm$ 1~K\cite{Mushnikov1999,Nohara2011,Aoki2011} by the sudden broadening of the step-like field dependencies of magnetization and thermoelectric power and the peak anomalies of the magnetoresistance and the Hall resistance\cite{Matsuda2000,Combier2013,Aoki2011,MATSUDA1999a,Kimura2015}. The MT in UTe$_2$ at low temperatures appears in a high magnetic field ($\approx$ 35~T\cite{Miyake2019,Knafo2019a,Niu2020}). Pulsed field magnetization data point to  $T_{CEP}$ = 11~K\cite{Miyake2019} whereas 7~K was reported from  magnetoresistance\cite{Knafo2019a} and the Hall resistance\cite{Niu2020} studies. This discrepancy in the reported $T_{CEP}$ values will be discussed below.

Already in the ground state, the IEMs appear on the verge of some instability that is accompanied by strong magnetic fluctuations. In UTe$_2$ magnetic fluctuations of U magnetic moments parallel to the $a$-axis were confirmed by inelastic neutron scattering\cite{Duan2020,Knafo2021a,Duan2021}, $^{125}$Te NMR\cite{Tokunaga2022}, and $\mu$SR\cite{Sundar2019} experiments. Ising ferromagnetic fluctuations are the key characteristics of UCoAl\cite{Mushnikov1999,Karube2012}. NMR spectroscopy is very sensitive to magnetic fluctuations. Field and temperature dependences of the Knight shift and nuclear spin-lattice and spin-spin relaxation rates 1/$T_1$ and 1/$T_2$, respectively, provided indications of the critical longitudinal magnetic fluctuations in the vicinity of CEP in UCoAl\cite{Nohara2011,Karube2012}. Tokunaga et al. conclude from the recent $^{125}$Te NMR experiment that UTe2 would be located on the paramagnetic side near an electronic phase boundary, where either magnetic or Fermi-surface instability would be the origin of the characteristic fluctuations\cite{Tokunaga2023}.

The magnetic interactions of ions in solids depend on their magnetic moment and neighborhood in a crystal lattice. Any displacement in the lattice leads to a renormalization of these interactions defining the coupling to the lattice degrees of freedom. This coupling results in static (shifts of ion positions) and dynamic (coupling to phonons) spin-lattice effects. Measurements of ultrasound attenuation and changes in ultrasound velocity due to ultrasound waves passing the sample of studied material provide important information on the character of the magnetic state and magnetic fluctuations\cite{Luthi2005}.

The strong magnetoelastic coupling indicated by the magnetostriction measured for UTe$_2$ and UCoAl suggests the possibility of studying the peculiar magnetic properties of the two IEMs by ultrasonic methods. Recently, Yoshizawa et al. showed that by using ultrasonic methods to measure the elastic properties of UCoAl in detail, it is possible to accurately identify the critical endpoint (CEP) within the $H-T$ phase space\cite{Yoshizawa2023}. They revealed pronounced softening of the longitudinal elastic stiffness $C_{33}$ in the vicinity of CEP that causes a corresponding reduction of sound velocity at CEP. The sound velocity drops also in water when it approaches the critical point\cite{Wagner2002}. This corroborates the suggested analogy between the CEP in the $H-T$ phase diagram of an IEM and CP in the $p-T$ diagram of water. Recent low-temperature ultrasound study of UTe$_2$ proposed the single-component character of its superconducting state \cite{Theuss2023} and showed the possibility of measuring the irregularly shaped samples by advanced resonant ultrasound spectroscopy \cite{Theuss2024}.    

We focus our main interest on the elastic phenomena in UTe$_2$ at various temperatures and magnetic fields in a broader neighborhood of suspected CEP in the $H-T$ phase space. To find common aspects corroborating the universality of elastic behavior of IEMs near CEP,  the results obtained on UTe$_2$ are discussed in comparison to the corresponding phenomena in UCoAl reported by Yoshizawa et al.\cite{Yoshizawa2023} and some data obtained by us on this material.

\section{\label{sec:Experimental}Methods}

High-quality single crystals of UTe$2$ were synthesized using the molten salt flux (MSF) technique\cite{Sakai2022}, yielding crystals of exceptional quality as evidenced by the highest superconducting transition temperature ($T_{SC}$ = 2.1~K) and residual resistivity ratio (RRR) values (up to 1000) reported for this material. These crystals originated from the same batch as those used in studies revealing quantum oscillations and expanding the understanding of high-field and high-pressure properties\cite{Eaton2024,Weinberger2023,Wu2024,Weinberger2024,Wu2024b,Wu2023}. Neither EDX chemical analysis, nor XRD structure analysis, nor detailed magnetization measurements indicated any magnetic impurity in the crystals. The needle-shaped crystals, measuring 5-7~mm along the $a$-axis, featured naturally flat and smooth surfaces perpendicular to the $c$-axis. Prior to ultrasonic investigations, the samples were checked by high-field magnetostriction measurements. The typical magnetostriction data measured on our UTe$_2$ crystals displayed in Fig. S1 of Supplementary Material\cite{supp} agree well with the literature\cite{Miyake2022}. 

UTe$_2$ has orthorhombic symmetry (space group $Immm$, \#71, $D_{2h}^{25}$). The single crystals were large enough to be cut and polished perpendicular to the $a$- and $c$-axis. We investigated 4 independent diagonal components of the elasticity tensor which are for the orthorhombic structure of UTe$_2$ described in Table~\ref{tab:Cij_UTe2}, where \textbf{\textit{q}} is the sound wave vector, \textbf{\textit{u}} is the displacement vector, \textbf{\textit{H}} is the magnetic field vector.
\begin{ruledtabular}
\begin{table}
    \centering
    \begin{tabular}{ccccc}
        Elastic mode & \textbf{\textit{q}} & \textbf{\textit{u}} & \textbf{\textit{H}} & 2 K (GPa)\\
        \colrule
        $C_{33}$ & [001] & [001] & [010] & 74(2)\\
        $C_{44}$ & [001] & [010] & [010] & 15(1)\\
        $C_{55}$ & [100] & [001] & [010] & 42(2)\\
        $C_{66}$ & [100] & [010] & [010] & 58(3)\\
    \end{tabular}
    \caption{Description of studied elastic modes in UTe$_2$. \textbf{\textit{q}} – direction of propagation vector of ultrasound, \textbf{\textit{u}} – direction of displacement vector, \textbf{\textit{H}}direction of magnetic field vector. Absolute value of the elastic constants was determined at 2~K}
    \label{tab:Cij_UTe2}
\end{table}
\end{ruledtabular}
The elastic properties under pulsed magnetic fields were examined at the Dresden High Magnetic Field Laboratory using a phase-sensitive detection technique\cite{Kohama2022}, with an ultrasound frequency of approximately 15~MHz for all modes and the magnetic field applied along the $b$-axis.

UCoAl has hexagonal symmetry (space group $P\overline{6}2m$, \#189, $D_{3h}^{3}$). The UCoAl single crystal was grown by a modified Czochralski method using the tri-arc furnace and consequently cut and polished with planes perpendicular to the $a-$ and $c-$axis, respectively. The ultrasound experiments were performed in transmission mode using the ultrasonic option of PPMS (Quantum Design) with the LiNbO$_3$ transducers glued by \textit{Thiokol LP032} to the sample for good acoustic contact. The measurements were performed on fundamental frequency for both longitudinal (33~MHz) and transversal (18~MHz) transducers. Measurements of the sound velocity at various temperatures and magnetic fields were done for the longitudinal $C_{33}$  and transversal $C_{44}$ acoustic modes as is shown in Table~\ref{tab:Cij_UCoAl}. 

The change in frequency ($\Delta f$) observed during the experiments can be converted to a change in velocity ($\Delta v$), with magnetostriction corrections ($\Delta L$). 
\begin{equation}
\frac{\Delta v}{v}=\frac{\Delta f}{f}+\frac{\Delta L}{L}
\label{eq:delta}
\end{equation}
The magnetostriction effects at the MT in both UTe$_2$ and UCoAl exhibit a magnitude on the order of 10$^{-4}$ (Fig. S1 of Supplementary Material\cite{supp} and Ref.\cite{Miyake2022} and \cite{HONDA2000}). Notably, this length change is at least one order of magnitude smaller than the observed frequency change.

The corresponding elastic constants ($C_{ij}$) are directly related to the sound velocity ($v_{ij}$), as

\begin{equation}
C_{ij} = \rho v_{ij}^2
\label{eq:Cij}
\end{equation}

where $\rho$ is the mass density. We used Eq.~\ref{eq:Cij} to calculate the absolute values of $C_{ij}$ at 2~K and zero magnetic fields for UTe$_2$ listed in Table~\ref{tab:Cij_UTe2} and for UCoAl listed in Table~\ref{tab:Cij_UCoAl}.  

\begin{ruledtabular}
\begin{table}
    \centering
    \begin{tabular}{ccccc}
        Elastic mode & \textbf{\textit{q}} & \textbf{\textit{u}} & \textbf{\textit{H}} & 2 K (GPa)\\
        \colrule
        $C_{33}$ & [001] & [001] & [001] & 206(14)\\
        $C_{44}$ & [001] & [100] & [001] & 245(17)\\
    \end{tabular}
    \caption{Description of studied elastic modes in UCoAl. \textbf{\textit{q}} – direction of propagation vector of ultrasound, \textbf{\textit{u}} – direction of displacement vector, \textbf{\textit{H}}direction of magnetic field vector.}
    \label{tab:Cij_UCoAl}
\end{table}
\end{ruledtabular}

\section{\label{sec:Results}Results and discussion}

The magnetoelastic coupling mediates a significant influence of magnetic fluctuations on the crystal lattice that may lead to lattice instabilities often connected with the softening of elastic modes. In this study, we explore these phenomena in UTe$_2$ by examining the elastic constants across various temperatures and magnetic fields near the MT using ultrasonic methods. It is important to note that Ushida et al.\cite{Ushida2023} have documented the temperature dependencies of $C_{33}$, $C_{44}$, and $C_{55}$, demonstrating a softening (hardening) of the $C_{55}$ ($C_{33}$) modes as the temperature decreases below 40~K. 

The principal finding of our study is the clear identification of the critical endpoint (CEP) within the magnetic field ($H$) and temperature ($T$) phase space, marked by a pronounced decrease in ultrasound velocity at a specific magnetic field strength, $H_{CEP}$, under constant temperature, $T_{CEP}$. This phenomenon is closely linked to the notable softening of elastic modes. We observed a minimum in the sound velocity in intermetallic UTe$_2$, a behavior that mirrors findings in UCoAl by Yoshizawa et al.\cite{Yoshizawa2023}. This similarity extends to the behavior of liquids at their critical point within the pressure-temperature ($p-T$) phase diagram, where sound velocity also reaches a minimum\cite{Fisher1990,Stishov2020}. This parallel suggests a universal aspect of phase transitions, highlighting the minimum in sound velocity as a characteristic feature near CEPs across different materials.

\begin{figure}
    \includegraphics[width=0.49\textwidth]{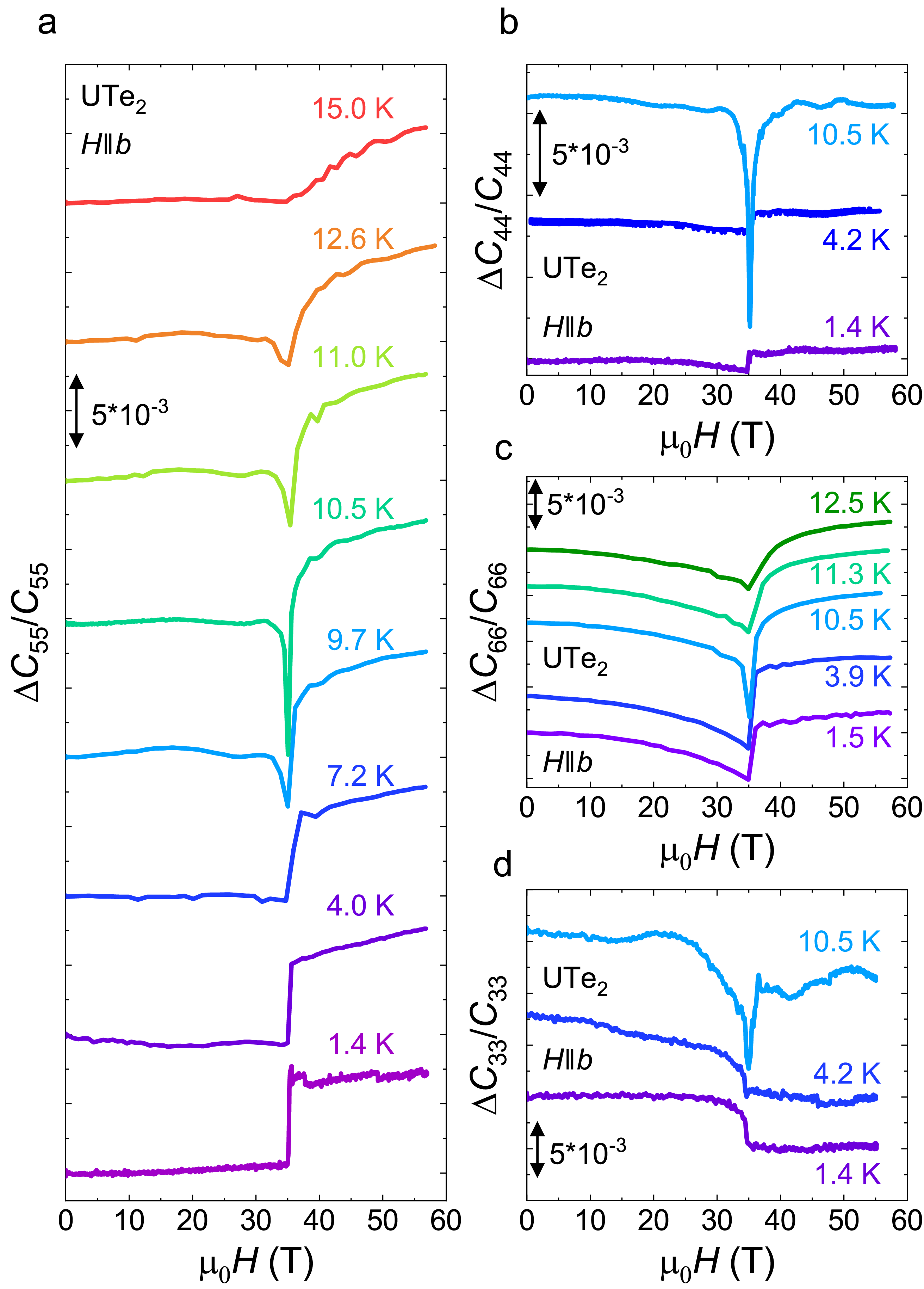} 
    \caption{\label{fig2}Relative change of the elastic constant $C_{55}$ (a), $C_{44}$ (b), $C_{66}$ (c) and $C_{33}$ (d) of UTe$_2$ versus magnetic field applied along the $b$-axis at several temperatures. The curves are shifted vertically for clarity} 
\end{figure}

\begin{figure}
    \includegraphics[width=0.49\textwidth]{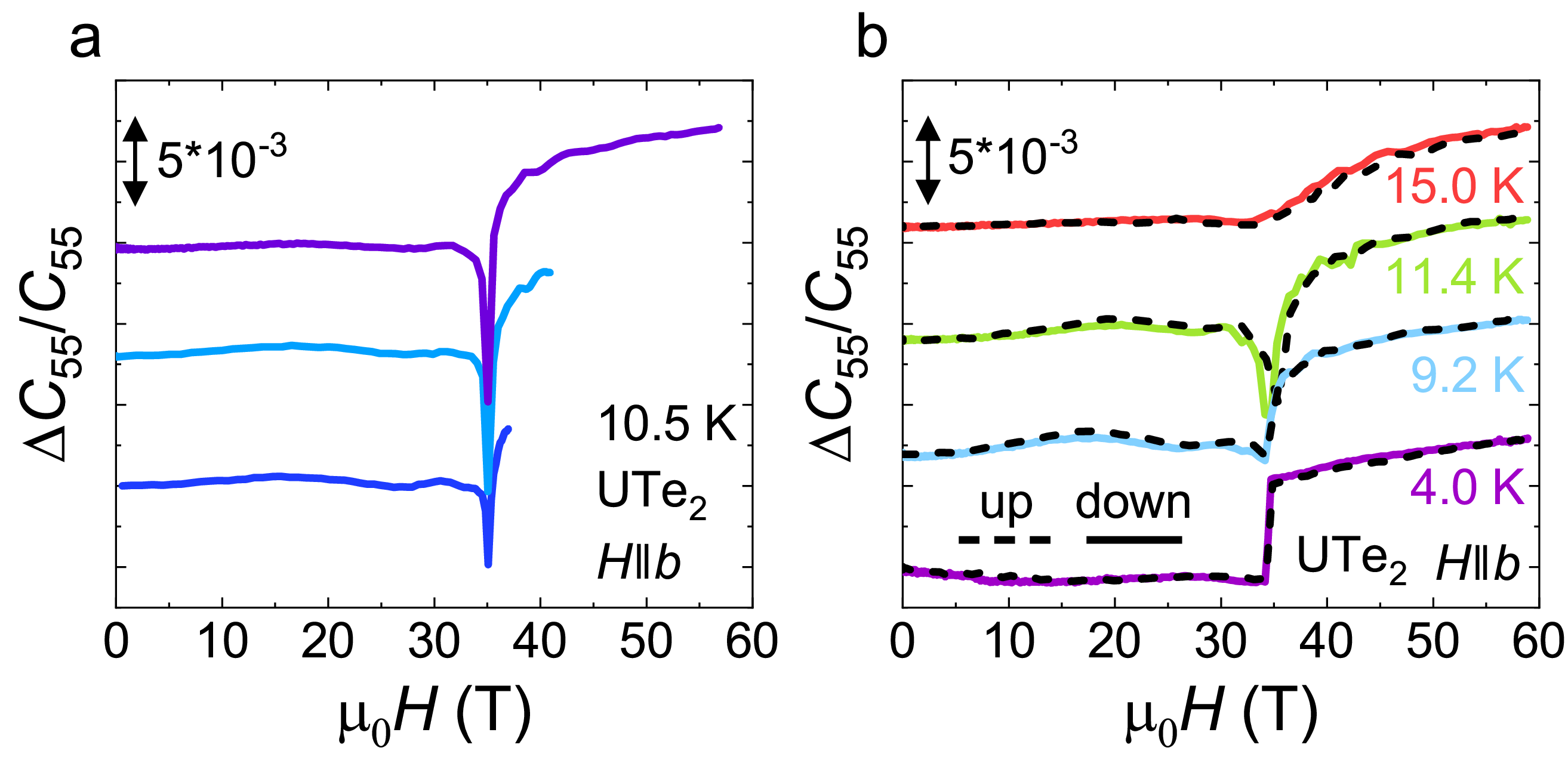} 
    \caption{\label{fig3}Relative change of the elastic constant $C_{55}$ of UTe$_2$ versus magnetic field applied along the $b$-axis (a)  at 10.5~K for different sweep rates due to the maximum magnetic fields, (b) at several temperatures showing the field sweeps up and down. The curves are shifted vertically for clarity} 
\end{figure}

Fig. ~\ref{fig2} shows our data of isothermal magnetic field dependences of the relative change of transversal $C_{44}$, $C_{55}$, $C_{66}$ modes, and longitudinal $C_{33}$ mode of UTe$_2$ determined at selected temperatures through the measured changes of ultrasound velocity corrected for magnetostriction. Among these, the $C_{55}$ mode exhibited a significantly clearer signal compared to the other three modes. Below 9~K, these modes show little to no magnetic field dependence, persisting up to and beyond a critical field strength of $\mu_0H_c\approx$ 35~T. At this juncture, UTe$_2$ undergoes a discontinuous first-order MT from the paramagnetic ground state to a polarized paramagnetic state at $H_c$, with the $C_{55}$ mode experiencing an abrupt increase of $\approx 0.7\%$. This behavior is reminiscent of certain antiferromagnets below the tricritical point, such as UIrSi$_3$\cite{Haidamak2022}, which exhibits a discontinuous first-order MT from an antiferromagnetic state to a polarized paramagnetic state at $H_c$, resulting in a sudden change of some diagonal elastic constants $C_{ij}$. At sufficiently low temperatures ($\leq$ 9~K in UTe$_2$), the acoustic response to the magnetic field increasing to $H_c$ is anticipated to be nearly independent of the field. However, at higher temperatures, a pronounced V-shaped dip emerges at $H_c$, reaching its deepest point of about 1\% at 10.5~K. This V-shaped anomaly gradually lessens with increasing temperature and vanishes around 15~K. The peak of this anomaly is consistently observed across different sweep rates and maximum applied field pulses, as shown in Figure ~\ref{fig3}a.

 We observed the sharp V-shape anomaly also for other elastic modes, $C_{44}$ (Fig.~\ref{fig2}b), $C_{66}$ (Fig.~\ref{fig2}c), and $C_{33}$ (Fig.~\ref{fig2}d). This anomaly consistently appears at 35~T and reaches its peak at 10.5~K across all four modes, with the magnitude ranging from 0.5\% to 1.5\%. The presence of this anomaly in the $C_{ij}$ versus $H$ during the midst of the metamagnetic transition distinctly marks $T_{CEP}$ at 10.5~K., where it is most pronounced due to the interaction between phonons and critical magnetic fluctuations in UTe$_2$. The identified $T_{CEP}$ closely aligns with the 11~K temperature previously reported by Miyake et al.\cite{Miyake2019} as $T_{CEP}$, characterized by a sudden shift in the sharpness of the $\partial M/\partial T$ versus $H$ peak from sharp at 10~K to more moderate at 11~K. A closer inspection of magnetoresistance data of Knafo et al.\cite{Knafo2019a} (that reports $T_{CEP}$ = 7~K) reveals that the determination of $T_{CEP}$  depends on the interpretation of the pronounced magnetoresistance maximum and also the curve measured at 10.5~K might be considered as a representative of CEP. Additionally, certain findings by Niu et al.\cite{Niu2020} might be interpreted as pointing to a $T_{CEP}$ around 10~K, such as the maximum in the $R_H$ versus $H$ curve for 10~K and the phase diagram analysis. 

The variation in reported CEP temperatures across studies may also be influenced by the magnetocaloric effect (MCE). Notably, a positive MCE has been observed in both field-up and field-down sweeps\cite{Schonemann2023}. The differences in sample sizes and their thermal coupling during various pulse field measurements could result in variations in adiabatic conditions. To explore this, we closely examined the data from both field-up and field-down sweeps. As shown in Fig. ~\ref{fig3}b, the congruence of the field-up and field-down data suggests that the MCE has a minimal impact on our ultrasound experiments, affirming our identification of $T_{CEP}$ at 10.5~K as accurate. In this context, we believe that also the magnetotransport data \cite{Knafo2019a,Niu2020} can be interpreted in terms of  $T_{CEP}$  near 10~K. 

Resembling features were observed by ultrasonic measurements of CeRu$_2$Si$_2$ at low temperatures\cite{Kouroudis1987,YANAGISAWA2002,Yanagisawa2002a,Bruls1990}. However, these observations do not confirm the presence of a metamagnetic CEP in this material at finite temperatures. Instead, a continuous softening of the material is observed as the temperature decreases, reaching down to the lowest measured temperature of 20 mK. This is in line with the fact that CeRu$_2$Si$_2$ appears in proximity to the magnetic quantum critical point but no metamagnetic transition is most likely realized in the low-temperature limit\cite{FLOUQUET2002} similar to the case of Sr$_3$Ru$_2$O$_7$\cite{Perry2001,Grigera2001,Ronning2006}. The observed lattice collapse can be understood due to the effect of critical magnetic fluctuations at a crossover between the ground state characterized by incommensurate antiferromagnetic correlations and the high-field polarized state. The critical magnetic fluctuations act on the crystal lattice via strong spin-phonon coupling. Similarly, UPt$_3$ does not undergo a true metamagnetic transition even at the lowest experimental temperatures\cite{Feller2000,Aeppli1987}.  Other Ce-based compounds, such as CeCu$_3$\cite{Goto1988} and CeNiSn\cite{HOLTMEIER1997}, also display magnetoacoustic anomalies that accompany pseudo-metamagnetic behavior, further illustrating the complex interplay between magnetic fluctuations and acoustic properties in these materials.

UCoAl has attracted great interest since the 1990s. As mentioned above, various studies have comprehensively characterized the metamagnetic behavior of this compound by measuring, the magnetic, thermal, transport properties and magnetostriction\cite{Matsuda2000,Combier2013,Aoki2011,MATSUDA1999a,Matsuda1999,Kimura2015} as well as NMR and NQR spectra\cite{Nohara2011,Karube2012}. The rich information about the magnetic phase diagram including the critical fluctuations in the vicinity of the critical end point motivated us to compare our findings for UTe$_2$ with results for UCoAl. The moderate values of characteristic temperatures and magnetic field in the UCoAl case enabled us to perform an additional ultrasonic study focused on the lattice response in the vicinity of the critical end point. As already mentioned in the Introduction most important information about CEP in UCoAl has been brought by Yoshizawa et al. by results of detailed ultrasound investigations, careful data analysis, and rich discussion of the complexity of the physics of this IEM\cite{Yoshizawa2023}.

\begin{figure}
    \includegraphics[width=0.48\textwidth]{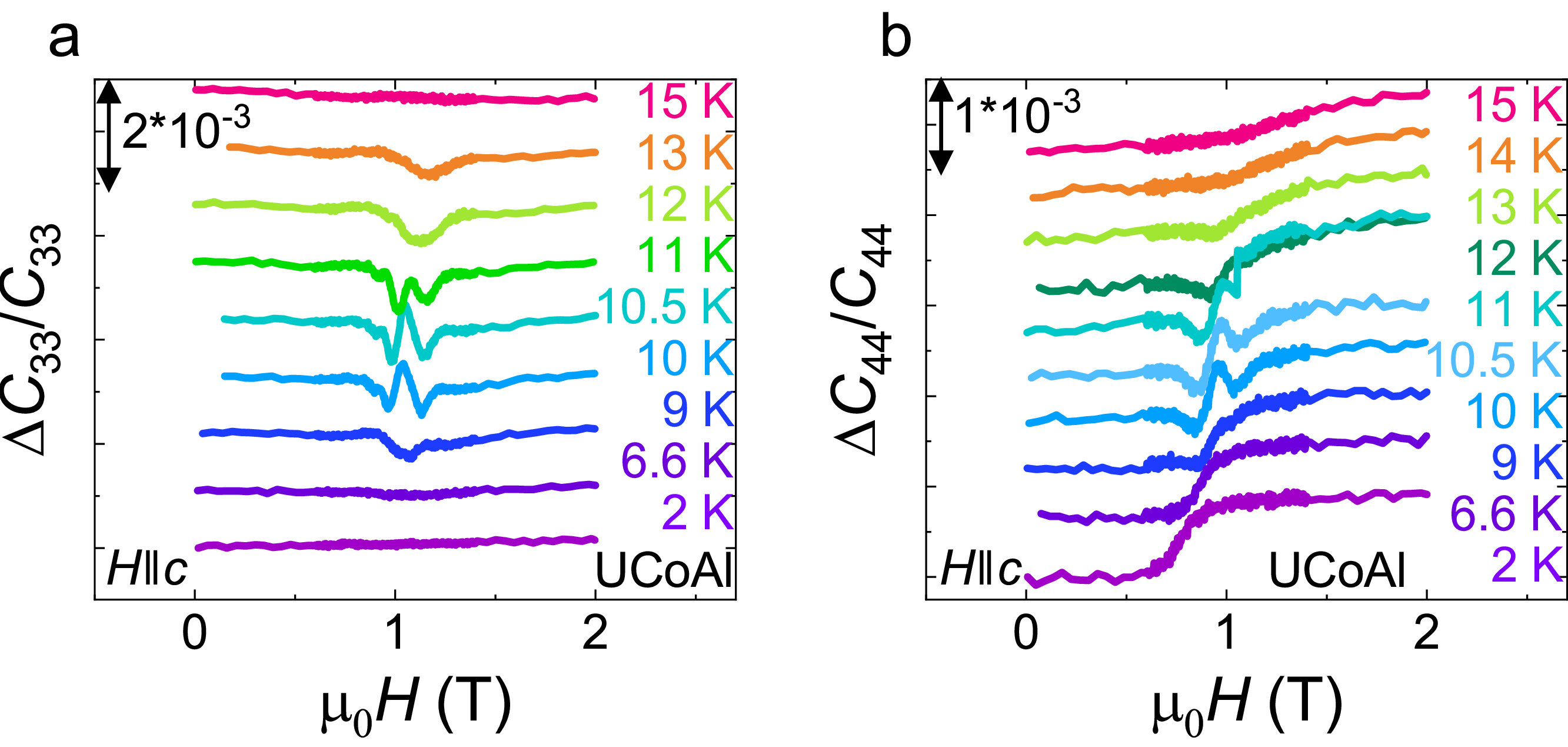} 
    \caption{\label{fig4}Relative change of the elastic constant $C_{33}$ (a) and $C_{44}$ (b) of UCoAl at several temperatures versus magnetic field applied along the $c$-axis. The ultrasonic frequency was 33 MHz (a) and 16.7 MHz (b). The curves are shifted vertically for clarity} 
\end{figure}

Our study uses UCoAl as a paradigm to showcase the universal behavior of IEMs at the CEP. Figure~\ref{fig4}a illustrates the behavior of the $C_{33}$ mode versus the magnetic field applied along the $c$-axis. At temperatures below 7~K and above 14~K, the field dependence is weak and nearly indistinct. However, between 10 and 11~K, a pronounced double anomaly (W-shape) emerges, peaking at a change of approximately 0.2\% at 10.5~K. This W-shape anomaly fades beyond 11~K and disappears above 13~K. At 9~K and between 12 to 13~K, a singular V-shaped minimum is noticeable. The $C_{44}$ versus $H$ plots, as shown in Figure ~\ref{fig4}b, exhibit a similar pattern but are influenced by an underlying signal that correlates with the magnetization increase due to the metamagnetic transition at temperatures up to 10.5~K and the crossover at higher temperatures. The observed W-shape may be suspected to be the result of a two-step metamagnetic transition. To check for this possibility we measured a series of representative magnetization isotherms of the same crystal used for acoustic study. The uniformly S-shaped curves of these isotherms (see Fig. S2\cite{supp}) dispel any doubts about a two-step transition, indicating an intrinsic characteristic of the material.

\begin{figure}
    \includegraphics[width=0.48\textwidth]{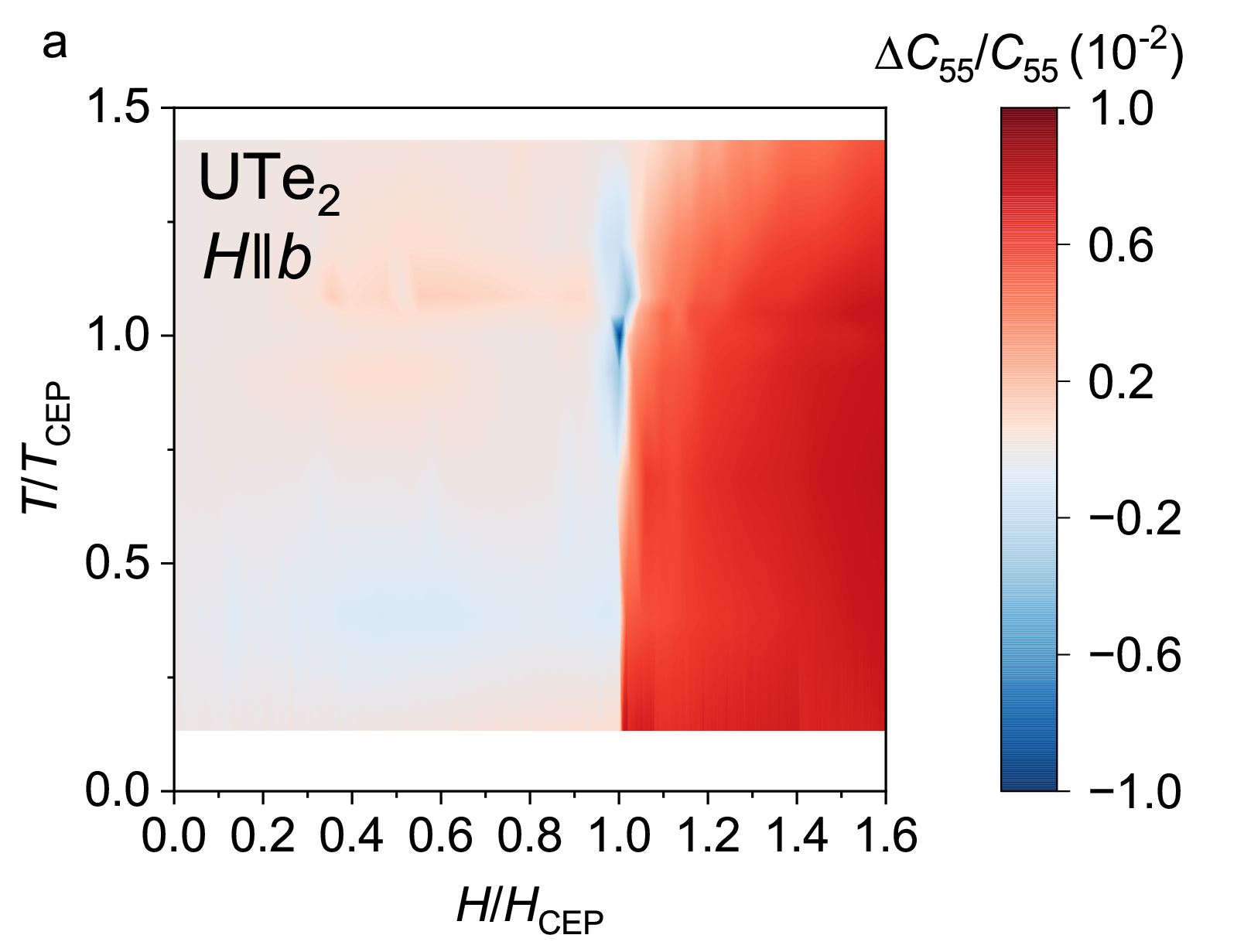}
    \includegraphics[width=0.48\textwidth]{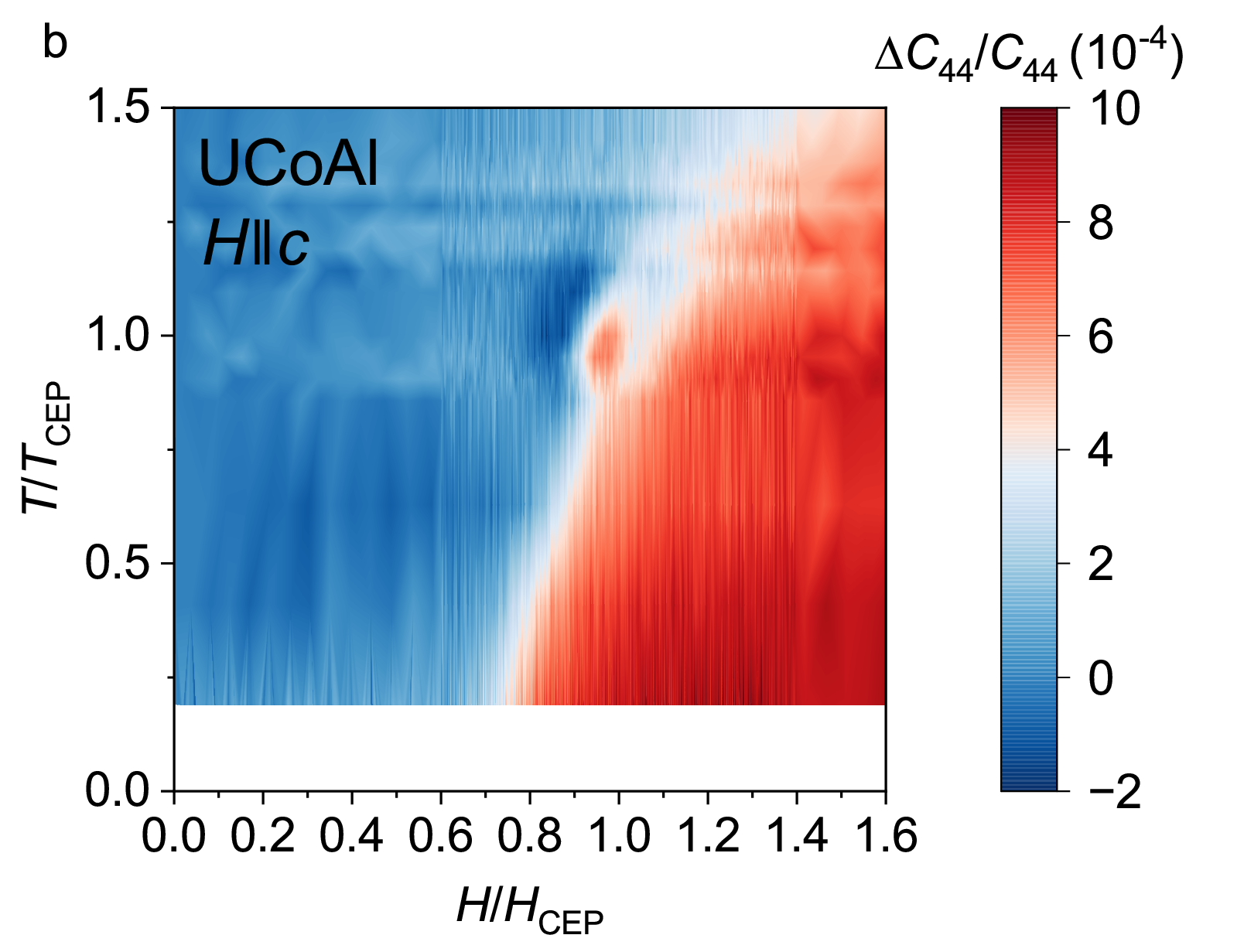} 
    \caption{\label{fig5}Colored contour plots of relative elastic constants $C_{55}$ of UTe$_2$ (a) and $C_{44}$ of UCoAl (b) in the $T/T_{CEP}$ - $H/H_{CEP}$ phase space.} 
\end{figure}

The evolution of the sharp W-shape anomaly in the isotherms $C_{ij}$  vs. $H$  measured on the hexagonal UCoAl crystal for $H\parallel c$ with temperature is similar to the development of the V-shape anomaly for the orthorhombic UTe$_2$ crystal for $H\parallel b$. Both anomalies reach their peak at the same temperature of 10.5~K, which is indicative of the CEP temperature ($T_{CEP}$). This parallel is further illustrated by the colored contour plots for $C_{55}$ in UTe$_2$ and $C_{44}$ in UCoAl in Fig.~\ref{fig5} (as well as $C_{33}$ in UCoAl in Fig. S3\cite{supp}), where an isolated spot—signifying a minimum or maximum in $C_{ij}$—is located at coordinates (1.0,1.0).

The distinction between the V-shape in UTe$_2$ and the W-shape in UCoAl is attributed to ultrasonic dispersion, which varies with frequency. This phenomenon is linked to the critical divergence of relaxation times near the metamagnetic transition at CEP, a subject thoroughly explored for UCoAl by Yoshizawa et al.\cite{Yoshizawa2023}. Conducting measurements at various frequencies allows for the observation of critical slowing down of fluctuations, where relaxation times are at their peak. At sufficiently low frequency a single V-shape is observed. The field (or temperature) dependence of elastic constant shows some structure (more than one simple V) when measured at higher frequencies becoming gradually more complex. In our experiments on UCoAl, we were able to receive a reasonable sensitivity for frequencies only down to 33~MHz and ended with the W-shape anomaly at $T_{CEP}$. 

\section{\label{sec:Conclusions}Conclusions}

In our study, we conducted ultrasound velocity measurements on UTe$_2$ and UCoAl, examining how their crystal lattices respond across various temperatures and magnetic fields. These measurements spanned multiple elastic modes, from which we derived the corresponding elastic constants. Our ultrasonic investigation sought to understand the behavior of the crystal lattice of IEMs near the first-order transition line in the $H-T$ phase diagram, which terminates at CEP—analogous to the critical point in the gas-liquid $p-T$ phase diagram. We discovered that either a sharp V-shape (in UTe$_2$) or W-shape (in UCoAl) anomaly appears at $H_c$ of MT in the isothermal field dependence of the elastic constant.  This anomaly peaks at the CEP characterized by $\mu_0H_c\approx$  35~T for UTe$_2$ and 1~T for UCoAl and $T_{CEP}$ = 10.5~K. Notably, the W-shape anomaly in UCoAl changes to a V-shape upon reducing the frequency. These anomalies near CEP highlight the lattice's instability triggered by critical magnetic fluctuations through strong magnetoelastic coupling, with the fluctuations reaching a supercritical state at CEP.

Although the paramagnetic ground state, lattice symmetry, and the expected process of the metamagnetic transition of UTe$_2$ and UCoAl are significantly different the evolutions of anomalies in elastic properties in the vicinity of CEP have some common aspects which may be universal for other IEMs. The identical $T_{CEP}$ values for both compounds are likely coincidental, yet the significant difference in their critical field ($H_c$) values—spanning one and a half orders of magnitude—merits further examination. UCoAl, characterized as an Ising metamagnet with uniaxially anisotropic magnetic moments and fluctuations, maintains its uniaxial symmetry at $T_{CEP}$.  The situation in UTe$_2$  is far not as straightforward as in UCoAl. The orthorhombic magnetic anisotropy and complex fluctuations due to the metamagnetic transition from the $a$-axis antiferromagnetic correlations in the ground state to the $b$-axis polarized moments in the high-field state will hamper the interpretation of ultrasonic data at CEP. A proper understanding of the lattice instabilities at CEP cannot be achieved without a fundamental theoretical background. We hope that these interesting phenomena will stimulate research efforts by theorists in studies of magnetoelastic coupling between magnetic and elastic fluctuations.

\begin{acknowledgments}
The UTe$_2$  and UCoAl crystals were grown and characterized and ultrasonic measurements on UCoAl were performed in MGML (\url{https://mgml.eu/}), which is supported within the program of Czech Research Infrastructures (project no. LM2023065). We acknowledge the support of the High Magnetic Field Laboratory (HLD) at Helmholtz-Zentrum Dresden-Rossendorf (HZDR), a member of the European Magnetic Field Laboratory (EMFL), the Deutsche Forschungsgemeinschaft (DFG) through SFB 1143,  and financial support by the Czech Science Foundation (GACR), project No. 22-22322S. The work of T.Y. was supported by JSPS KAKENHI Grant Numbers JP21KK0046, JP22K03501, JP22H04933, and JP23H04868.” A part of this work was also supported by the JAEA REIMEI Research Program.
\end{acknowledgments}

%

\end{document}